\documentclass[%
reprint,
 amsmath,amssymb,superscriptaddress,
longbibliography,
prl
]{revtex4-1}
\usepackage{xparse}

\usepackage{appendix}
\usepackage{graphicx}
\usepackage{dcolumn}
\usepackage{bm}
\usepackage{hyperref}

\newcommand{\Tr}{\text{Tr}}
\usepackage{slashed}
\renewcommand{\vec}[1]{\mathbf{#1}}
\newcommand{\Lim}[1]{\raisebox{0.5ex}{\scalebox{0.8}{$\displaystyle \lim_{#1}\;$}}}

\usepackage{bbold}

\begin{document}
\title{Chiral Anomaly in interacting Condensed Matter Systems}

	\author{Colin Rylands}
\affiliation{Joint Quantum Institute and
 Condensed Matter Theory Center, University of Maryland, College Park, MD 20742, USA}
\author{Alireza Parhizkar}
\affiliation{Joint Quantum Institute and
	Condensed Matter Theory Center, University of Maryland, College Park, MD 20742, USA}
 	\author{Anton A. Burkov}
\affiliation{Department of Physics and Astronomy, University of Waterloo, Waterloo, Ontario N2L 3G1, Canada}
\affiliation{Perimeter Institute for Theoretical Physics, Waterloo, Ontario N2L 2Y5, Canada}
\author{Victor Galitski}
\affiliation{Joint Quantum Institute and
	Condensed Matter Theory Center, University of Maryland, College Park, MD 20742, USA}

\date{
    \today
}

\begin{abstract}
The chiral anomaly is a fundamental quantum mechanical phenomenon which is of great importance to both particle physics and condensed matter physics alike. In the context of QED it manifests as the breaking of chiral symmetry in the presence of electromagnetic fields. It is also known that anomalous chiral symmetry breaking can occur through interactions alone, as is the case for interacting one dimensional systems. In this paper we investigate the interplay between these two modes of anomalous chiral symmetry breaking in the context of interacting Weyl semimetals. Using Fujikawa’s path integral method we show that the chiral charge continuity equation is modified by the presence of interactions
which can be viewed as including the effect of the electric and magnetic fields generated by the interacting quantum matter.
This can be understood further using dimensional reduction and a Luttinger liquid description of the lowest Landau level. These effects manifest themselves in the non-linear response of the system. In particular we find an interaction dependent density response due to a change in the magnetic field as well as a contribution to the non-equilibrium and inhomogeneous anomalous Hall response while preserving its equilibrium value.
\end{abstract}
\date{\today}

\maketitle


\textit{Introduction}---Modern condensed matter physics has benefited greatly from concepts originally introduced in the context of high energy physics. One such concept is the chiral anomaly; the breaking of classical chiral symmetry in a quantum theory~\cite{Adler, BellJackiw}. Within QED it arises through the need to regularize certain loop diagrams  which contain differences of linearly divergent integrals. The appropriate regularization can either preserve charge conservation symmetry, chiral symmetry or some combination of the two but not both. On physical grounds, the first of these is chosen, which brings about a source term for the divergence of the chiral current, $j_5^\mu$, whenever electric and magnetic fields are not orthogonal, \begin{equation}\label{Anomaly}
    \partial_\mu j^\mu_5=\frac{e^2}{2\pi^2}\vec{E}\cdot\vec{B}.
\end{equation}Here $\vec{E}$ and $\vec{B}$ are the electric and magnetic fields and we have set $c=\hbar=1$.  This expression, although derived from a single triangle diagram in perturbation theory was shown to obey non-renormalization theorems; higher order terms cannot modify the form of this equation and are accounted for by replacing the bare fields and charge with their renormalized values~\cite{AdlerBardeen}.  Later, this was reinforced when it was discovered that the chiral anomaly manifests in the path integral formalism through the non-invariance of the measure under a chiral symmetry transformation~\cite{Fujikawa, FujikawaErrata,Fujikawa2004Book}.  

 The chiral anomaly is present  for all odd spatial dimensions~\cite{FramptonKephart, FramptonKephartErrata, ZuminoYongShiZee} but is particularly important in one spatial dimension where it is crucial for the proper treatment of interacting fermionic theories through bosonization~\cite{Naon, LeeChen}. A prominent feature therein is that chiral symmetry breaking can occur due to the presence of interactions even when electromagnetic fields are absent. Indeed, it is well known, although perhaps not expressed in this way, that the chiral charge conservation equation for interacting fermions is~\cite{GiamarchiBook, GogolinNerseyanTsvelikBook}
 \begin{eqnarray}\label{1danomaly}
 \partial_\mu j_5^\mu=\frac{\lambda^2}{2\pi}\partial_1 j_5^1 ,
 \end{eqnarray}
 where $\lambda^2/2$ is the strength of the density-density interactions and the index $1$ refers to the spatial direction. By writing the expression in this form we have separated out the part which appears due to the non-invariance of the path integral measure. If an electric field is present also, it will appear as an additional $eE/\pi$ term on the right hand side~\cite{NielsenNinomiya}.

Chiral symmetry, is an emergent low energy property in condensed matter systems, appearing due to an even number of chiral modes crossing the Fermi surface which are actually part of the same band. In this respect, the anomaly can be understood in non-interacting systems via the pumping of chiral charge through the bottom of the band from one node to another~\cite{NielsenNinomiya}. Despite not being a fundamental symmetry, it is intimately related to many key concepts including the quantized Hall conductance, e.g. through Laughlins's argument~\cite{Laughlin}, and more recently the existence of topological metals such as the Weyl semimetal~\cite{WanTurnerVishwanathSavrasov,BurkovBalents, YangLuRab, XuWengWangDaiFang, HalaszBalents, Aji,WengFangZhongBernevigDai, Lv1, Lv2, Xu, Huang}. In this Letter  we examine the interplay between the two modes of chiral symmetry breaking expressed through \eqref{Anomaly} and \eqref{1danomaly} in the context of interacting condensed matter systems.  Specifically, we show that for short range interactions the anomaly can be written as 
\begin{eqnarray}\label{Result}
    \partial_\mu j^\mu_5=\frac{e^2}{2\pi^2}\tilde{\vec{E}}\cdot\tilde{\vec{B}},
\end{eqnarray}
where $\tilde{\vec{E}}$ and $\tilde{\vec{B}}$, defined below, contain the effect of both the electromagnetic fields in a manner similar to~\eqref{Anomaly} and the interactions through terms like in~\eqref{1danomaly}.

 The effect of interactions in Weyl semimetals has been considered previously using perturbative means~\cite{YongKi, GorbarMiranskyShovkovy, GorbarMiranskyShovkovySukhachov, MiranskyShovkovy, MaciejkoNandkishore}. In contrast, our work takes a non-perturbative approach and considers the interactions from the outset through the chiral anomaly itself. By utilizing \eqref{Result} we predict a number of new non-perturbative phenomena found beyond linear response which can be expected in interacting Weyl semimetals and attributed to the chiral anomaly. 


\textit{Model}---We consider a model of interacting Dirac fermions, $\psi$, in the presence of a constant background  magnetic field in $3+1$ dimensions. The action is $S=S_0+S_\text{int}$ with
\begin{eqnarray}
S_0=\int d^4x\,\bar{\psi}(x)\left[i\slashed{\partial}+e\slashed{A}\right]\psi(x) ,
\end{eqnarray}
where we have employed Dirac slash notation and $\bar{\psi}=\psi^\dag\gamma_0$. For later convenience we  split the gauge field, $A^\mu=A^\mu_{0}+\tilde{A}^\mu$, into a part describing the magnetic field pointing along the $\hat{z}$ direction, $A^\mu_{0}=xB_z \delta^\mu_2$ and a perturbation  around it, $\tilde{A}^\mu$. The magnetic field breaks the Lorentz invariance down to rotational invariance in the transverse plane spanned by the $\hat{x}$ and $\hat{y}$ directions  and  reduced $(1+1)$-d Lorentz symmetry in the longitudinal directions. The general short range current-current interaction is of the form
\begin{equation}
	S_\text{int} = -\frac{1}{2} \int d^4x \, \lambda^2_{\mu\nu} j^\mu(x) j^\nu(x) ,
\end{equation}
where $j^\mu(x)=\bar{\psi}(x)\gamma^\mu\psi(x)$ is the fermion current with $\lambda^2_{\mu\nu}=\lambda_{\mu\alpha}\lambda_\nu^{\ \alpha}$ being the interaction strength. Although certain results presented in this Letter are more general, we have restricted our focus to special cases of $\lambda^2_{\mu\nu}=\lambda^2\eta_{\mu\nu}$, which preserves Lorentz symmetry,  $\lambda^2_{\mu\nu}=\lambda^2_0\eta_{0\mu}\eta_{0\nu} + \lambda^2_3\eta_{3\mu}\eta_{3\nu}$ which preserves the reduced symmetries of our system if $\lambda^2_0=\lambda^2_3$ and which gives density-density interaction when $\lambda^2_3=0$~\footnote{$\eta_{\mu\nu}$ is the metric of the space-time which for our purposes here is considered to be flat. Also, $\delta^\mu_\nu$ is the Kronecker delta. Throughout the Letter we use Einstein's notations and sometimes represent the four-vector of current as $j^\mu \equiv (\rho,j^x,j^y,j^z)$ in Minkowski coordinates $x^\mu \equiv (t,x,y,z)$. }. Evidently, depending on the choice of $\lambda_{\mu\nu}$ some of the symmetries of the model may be broken, e.g. Lorentz invariance, but they do not break the classical chiral symmetry. These interactions are RG irrelevant and typically are not considered, however we will see that in the presence of the constant magnetic field, they should not be discounted. 

\textit{Chiral Anomaly \& Interactions}---To study the chiral anomaly in the presence of interactions we proceed using a generalization of Fujikawa's path integral method~\cite{Fujikawa, FujikawaErrata}. The  path integral is
\begin{equation} \label{HubStra}
    I =\int\mathcal{D}\left[\bar{\psi}\psi a_\mu\right]\exp{i \Big\{\int d^4x\,\bar{\psi}i\slashed{D}\psi+\frac{1}{2}a_\mu a^\mu}\Big\} ,
\end{equation}
where we have introduced the  Hubbard-Stratonovich field $a_\mu(x)$ which has been included in the generalized Dirac operator as $D_\mu=\partial_\mu -ieA_\mu - i\lambda_{\mu\nu}a^\nu$, and whose equation of motion reads $a_\mu= - \lambda_{\nu\mu}j^\nu$. Integration over the auxiliary $a_\mu$ field gives the original action $S=S_0 + S_\text{int}$ back. We now perform an infinitesimal chiral transformation, $\psi\to e^{i\theta(x)\gamma_5 }\psi, ~\bar{\psi}\to \bar{\psi} e^{i\theta(x) \gamma_5 }$  which results in a shift of the action,
\begin{eqnarray}
    S\to S+ \int d^4x \,\theta(x)\left[\partial_\mu j^\mu_5-\mathcal{A}_5(x)\right] ,
\end{eqnarray}
where  $j^\mu_5(x)=\bar{\psi}(x)\gamma^\mu\gamma_5 \psi(x)$ is the chiral current. The first term in the brackets arises from the classical shift of the action itself whereas the second is the anomalous term which is a result of the non-invariance of the measure. It takes the standard form $\mathcal{A}_5(x)=2\Tr{\left[\theta(x)\gamma_5\right]}$ or more explicitly
\begin{eqnarray}\label{Trace}
    \mathcal{A}_5(x)=2 \theta(x)\sum_n \varphi^\dagger_n(x)\gamma_5\,\varphi_n(x) ,
\end{eqnarray}
where $\varphi_n(x)$ are some orthonormal basis of wavefunctions used to expand the Grassmann variables $\psi(x)=\sum c_n\phi_n(x)$. In the absence of interactions the natural choice is to take these to be the eigenfunctions of  $\slashed{D}_0=\slashed{\partial}-ie\slashed{A}$ and regularize this divergent sum using the heat kernel method $\sum_n\to \lim_{M\to 0 }\sum_ne^{-\slashed{D}^2_0/M}$. Such a choice of basis has the crucial benefit of formally diagonalizing the action. This results in the familiar anomalous term $\mathcal{A}_5(x)=\theta(x) \frac{e^2}{16\pi^2}F_{\mu \nu}F_{\rho \sigma}\epsilon^{\mu\nu\rho\sigma}$ with $F_{\mu \nu}=\partial_\mu A_\nu-\partial_\nu A_\mu$ and $\epsilon^{\mu\nu\rho\sigma}$ is the Levi-Cevita symbol. The chiral anomaly equation~\eqref{Anomaly} then follows. Note that owing to the fact that $\{\gamma_5 ,\slashed{D}_0\}=0$ it is evident from~\eqref{Trace} that anomalous term is generated solely by the zero modes of the Dirac operator. 

In the presence of interactions we regularize the sum using the generalized Dirac operator, including the Hubbard-Stratonovich field  $\slashed{D}=\gamma^\mu(\partial_\mu -ieA_\mu -i \lambda_{\mu\nu}a^\nu)~$(For similar approaches see~[\citenum{RainesGalitski,ZyuzinBurkov}]). Following the same procedure we find, $\mathcal{A}_5(x)=\theta(x) \frac{1}{16\pi^2}\mathcal{F}_{\mu \nu}\mathcal{F}_{\rho\sigma}\epsilon^{\mu\nu\rho\sigma}$ where $\mathcal{F}_{\mu\nu}=\partial_\mu (eA_\nu+\lambda_{\nu\alpha}a^\alpha)-\partial_\nu (eA_\mu+\lambda_{\mu\beta}a^\beta)$ and after integrating over $a_\mu$ we find
\begin{eqnarray}\nonumber\label{InteractingAnomaly}
\partial_\mu j^\mu_5&=&\frac{e^2}{16\pi^2}F_{\mu \nu}F_{\rho \sigma}\epsilon^{\mu\nu\rho\sigma}
-\frac{e}{2\pi^2}\epsilon^{\mu\nu\rho\sigma}\lambda^2_{\sigma\alpha}\partial_\mu A_\nu\partial_\rho j^\alpha \\
&&+\frac{1}{4\pi^2}\epsilon^{\mu\nu\rho\sigma}\lambda^2_{\nu\alpha}\lambda^2_{\sigma\beta}\partial_\mu j^\alpha \partial_\rho j^\beta. 
\end{eqnarray}
We see that there are  terms depending only on the electromagnetic field, only on the presence of interactions and a mixed term requiring the presence of both. After defining
\begin{eqnarray}\label{D}
\tilde{E}_i&=&E_i-\frac{1}{e}\left[\lambda^2_{i\beta}\partial_0-\lambda^2_{0\beta}\partial_i\right]j^\beta,\\\label{H}
\tilde{B}_i&=&B_i-\frac{1}{2e}\epsilon_{ijk}\left[\lambda^2_{j\beta}\partial_k-\lambda^2_{k\beta}\partial_j\right]j^\beta,
\end{eqnarray}
equation \eqref{Result} is obtained. 

We could view this as a screening, by the interactions of the electric and magnetic fields which are responsible for the non-conservation of the chiral charge.  This can be seen more clearly by allowing the electromagnetic fields to be dynamical and, for simplicity, considering $\lambda^2_{\mu\nu}=\lambda^2\eta_{\mu0}\eta_{\nu0}$, i.e. density-density interactions. Upon treating the electromagnetic field in a semi-classical fashion through $ej^\nu=\partial_\mu F^{\nu\mu}$, we find that $\tilde{\vec{E}}=\vec{E}-\frac{\lambda^2}{e^2}\vec{\nabla}\left(\vec{\nabla}\cdot \vec{E}\right)$ and $\tilde{\vec{B}}=\vec{B}$.  Therefore the anomalous chiral symmetry breaking is generated not only by the background fields but also by the fluctuations induced by the interacting matter.

 \textit{Dimensional reduction to a Luttinger liquid}---
The chiral anomaly, in the free case, can be straightforwardly understood through dimensional reduction of the $(3+1)$-d system to the $(1+1)$-d linearly dispersing lowest Landau level (LLL) via a  magnetic field, $B_z$~\cite{NielsenNinomiya}.
We show now that one can also arrive at \eqref{Result} using dimensional reduction provided that the LLL is described by a Luttinger liquid. 

 Let us consider a  system that is homogeneous in the transverse directions along $\hat{x}$ and $\hat{y}$. In particular, the only external fields are in the longitudinal $\hat{z}$ direction and there are no currents which vary along $\hat{x}$ and $\hat{y}$. Our anomalous relation then reduces to
 \begin{equation}
\partial_\mu j^\mu_5=\frac{e^2}{8\pi^2}F_{\mu \nu}F_{\rho \sigma}\epsilon^{\mu\nu\rho\sigma}-\frac{e B_z}{2\pi^2}\lambda^2_{\sigma\alpha}\epsilon^{12\rho\sigma}  \partial_\rho j^\alpha.
\end{equation}
Assuming that the interacting system still forms Landau levels, the zero modes which are responsible for the anomaly  are present only on the LLL. As in the free case, the magnetic field  achieves a dimensional reduction from the $(3+1)$-d theory to the LLL which is effectively $(1+1)$-d. Within the LLL the following identity is valid $\epsilon^{12\rho\sigma} \gamma_\sigma=\gamma_5 \gamma^\rho$ and after some rearranging we arrive at
\begin{equation}\label{reducedAnomaly}
    \partial_\mu j^\mu_5= \frac{1}{1+n_0\lambda^2_{3}/\pi}\frac{e^2}{2\pi^2}E_zB_z -\frac{n_0\left(\lambda^2_{0} - \lambda^2_{3}\right)/\pi}{1+n_0\lambda^2_{3}/\pi} \partial_3 j^3_5 ,
\end{equation}
where  $n_0=\frac{eB_z}{2\pi}$. Here we have also specialized to the case where the interaction tensor is diagonal.  In deriving this equation we have assumed that Landau levels are formed in the interacting system or more precisely that there is a spin polarized LLL on which the anomaly is generated. We have made no assumptions on the nature of Landau levels or how they arise, only that they exist which seems a physically reasonable proposition especially in the limit of large background field.   In the opposite limit of zero background field \eqref{reducedAnomaly} reduces to the noninteracting result. 

The second term in \eqref{reducedAnomaly} is similar to \eqref{1danomaly} while the modification of the first has been discovered before in early studies of interacting $(1+1)$-d fermions~\cite{GeorgiRawls, SunSheng}. 
 To understand their appearance better we introduce the following action consisting of $N$ coupled $(1+1)$-d bosonic fields
\begin{eqnarray}\nonumber
    S&=&\sum_{j=1}^N\int \frac{d^2x}{2\pi} \Bigg\{\left[\partial_t\phi_j\right]^2+\left[\partial_x\phi_j\right]^2
    -e\left[\epsilon^{mn}A_{m}\partial_n\right]\phi_j \\
    &&+\sum_{j\leq k}\frac{\lambda^2_0}{\pi}[\partial_x\phi_j][\partial_x\phi_k]+\frac{\lambda^2_3}{\pi}[\partial_t\phi_j][\partial_t\phi_k]\Bigg\} ,
\end{eqnarray}
with $\epsilon^{mn}$ the $2$-d Levi-Cevita symbol. This is equivalent, through bosonization, to a system of $N$ flavors of interacting chiral fermions, $\chi^\dag_{\pm,j}=\sqrt{\rho_0}e^{i \left[\pm\phi_j-\int^t dt\partial_x\phi_j\right]}$ where $\rho_0$ is the background density~\cite{GiamarchiBook, GogolinNerseyanTsvelikBook}. The bosons are related to the fermionic charge and chiral charge density via $\sum_{\sigma=\pm}:\chi^\dag_{\sigma,j}\chi_{\sigma,j}:=-\partial_x\phi_j/\pi$  and $\sum_{\sigma=\pm}\sigma:\chi_{\sigma,j}^\dag\chi_{\sigma,j}:=\partial_t\phi_j/\pi$ with $:\,:$ indicating normal ordering.  

The model is flavor symmetric and accordingly both the interactions and the gauge field affect only the symmetric combination, $\phi_S=\frac{1}{\sqrt{N}}\sum_j\phi_j$. After a canonical transformation and retaining only the symmetric terms we arrive at the following action
\begin{eqnarray}\nonumber
S_S&=&\int \frac{d^2x}{2\pi} \left(1+\lambda_0^2N/\pi\right)[\partial_x\phi_S]^2+ \left(1+\lambda_3^2N/\pi\right)[\partial_t\phi_S]^2\\\label{SymmetricBoson}&&-2 \sqrt{N} eA_0\partial_x\phi_S+2 \sqrt{N}eA_3 \partial_t\phi_S.
\end{eqnarray}
Note that here the gauge field couples to the fermionic density rather than through minimal coupling with the symmetric boson, an important distinction which we comment on further below. The chiral anomaly is now manifest in the Euler-Lagrange equation for $\phi_S$. Calculating this we find  agreement with~\eqref{reducedAnomaly} provided one identifies the number of flavors with the Landau level degeneracy, $N=n_0=eB_z/2\pi$ as well as $j_5^0=\sum\partial_t\phi_j/\pi$ and $j_5^3=\sum\partial_x\phi_j/\pi$ which follows from the properties of $\gamma^\mu$ in $(1+1)-$d. 

Our path integral calculation is therefore consistent with a description of the LLL as a Luttinger liquid. A Luttinger liquid approach has also been adopted in~[\citenum{ZhangNagaosa}] to investigate the effect of disorder which we shall not consider here. The Luttinger liquid consists of a pair of interacting chiral fermions $\chi^\dag_{\pm,S}=\sqrt{\rho_0}e^{i \left[\pm\phi_S-\int^t dt\partial_x\phi_S\right]}$ formed from the symmetric boson which couple to the gauge field and the decoupled non-symmetric fields which play no role. The excitations of the LLL are still chiral but are distinct from these bare fermions and are created by $\Psi^\dag_{\pm}=\sqrt{\rho_0}e^{i \left[\pm\sqrt{1+\lambda_0^2N/\pi}\phi_S- \sqrt{1+\lambda_3^2N/\pi}\int^t dt\partial_x\phi_S\right]}$ which coincide with $\chi^\dag_{\pm,S}$ only when interactions are absent. In general these excitations  carry different electric and chiral charges from $\chi^\dag_{\pm,s}$ which can be seen through the coefficients of $\phi_S$ and $\int^t dt\partial_x\phi_S$ in the exponential. Had our gauge field coupled to these instead then we would find that the chiral anomaly equation was unmodified. 
A similar situation also arises when comparing conductances in one dimensional systems~\cite{AlekseevCheianovFrohlich}. 

As mentioned in the introduction, the chiral anomaly is related to Laughlin's argument for quantized Hall conductance~\cite{Laughlin}. Therein one can argue that the invariance of the Hall conductance to local interactions implies invariance of the chiral anomaly for the edge modes of Laughlin's cylinder and vice versa. We remark that our results are not in contradiction to this as our $(1+1)$-d chiral modes are not spatially separated as they are in Laughlin's argument. In order to see similar interaction effects as ours one would need to include non-local interactions between the edges.


\textit{Consequences for Weyl Semimetals }---We now turn our attention to the consequences of \eqref{Result} for interacting interacting condensed matter systems, in particular  Weyl semimetals. These are a recently discovered type of gapless topological matter possessing a number of distinctive features which arise due to the chiral anomaly including a large negative magnetoresistance~\cite{NielsenNinomiya, SonSpivak,Burkov, FukushimaKharzeevWarringa} and an anomalous Hall response~\cite{ZyuzinBurkov,ChenWuBurkov}. The low energy description of such systems is given by $S=S_0+S_b+S_\text{int}$ with $S_b = \int d^4x \, b_\mu j_5^\mu,$
where $ b_\mu$ separates the Weyl nodes in momentum and energy space. The effect of this term is most conveniently seen by performing a chiral rotation $\psi\to e^{ib_\mu x^\mu\gamma_5 }\psi, ~\bar{\psi}\to \bar{\psi} e^{i b_\mu x^\mu\gamma_5 }$  which removes $S_b$ at the cost of generating a Chern-Simons term, $S_{CS}$ due to the  chiral anomaly. In terms of the Hubbard-Stratonovich field this is
\begin{align}\label{ChernSimons}
S_{CS}=\int \frac{d^4x}{4\pi^2}\epsilon^{\nu\mu\rho\sigma}b_\mu \left[e A_\nu+\lambda_{\nu\alpha}a^\alpha\right]\partial_\rho \left[eA_\sigma+\lambda_{\sigma\beta}a^\beta\right].
\end{align} Then, following~[\citenum{ZyuzinBurkov}] we vary $S+S_{CS}$ with respect to $A_1$ to obtain the anomalous Hall current. Specializing to the case $b_\mu=b_z\delta^3_\mu$, $\lambda_{\mu\nu}=\lambda\eta_{\mu\nu}$ and after integrating over $a_\mu$ we find $j^x=\frac{eb_z}{2\pi^2}\tilde{E}^y$ or more expicitly
\begin{eqnarray}\label{HallResponse}
  j^x=\frac{eb_z}{2\pi^2}E^y-\frac{\lambda^2 b_z}{2\pi^2}\left[\partial_t j^y-\partial_{y}\rho\right],
\end{eqnarray}
with $E^y$ being the electric field along $\hat{y}$ and $\rho(x)=j^0(x)$. The first term here gives the quantum anomalous Hall current while the interaction dependent contribution vanishes in equilibrium. Thus, the interactions do not affect the equilibrium Hall current however they may contribute to the non-equilibrium or inhomogeneous response.  Combining~\eqref{HallResponse} with the corresponding expression for $j^y$ and switching to Fourier space we obtain the homogeneous finite frequency Hall conductivity expected from $S_{CS}$,
\begin{eqnarray}
 \sigma^{xy}(\omega)=\left[1+\left(\frac{\lambda^2b_z}{2\pi^2}\omega\right)^2\right]^{-1}\frac{e^2b_z}{2\pi^2}.
\end{eqnarray}
The effect of interactions can also be seen in the equilibrium density response to a change in the magnetic field,  $B_z\to B_z+ \delta B_z$. In the absence of any fields along the transverse components we may use \eqref{reducedAnomaly} as our anomalous relation. After subtracting the background density, the leading order density response is 
\begin{eqnarray}\label{ChiralSeparationEffect}
 \delta j^0=\frac{1}{1+\lambda^2\frac{e B_z}{2\pi^2}} \frac{eb_z}{2\pi^2}\delta B_z.
\end{eqnarray}
Due to the dimensional reduction, the density is equivalent to a chiral current in the longitudinal direction, $\left<j^0\right>=\left<j^3_5\right>$ and so~\eqref{ChiralSeparationEffect} can be viewed as the generation of a chiral current in response to a change in the magnetic field which is known as the chiral separation effect (CSE)~\cite{Vilenkin, MetlitskiZhitnitsky, NewmanSon}.

 \textit{Photon Action}---As was pointed out in~\cite{ChenWuBurkov} the Chern-Simons term obtained via chiral transformation requires some subtle interpretation if it is to describe a Weyl semimetal.
The appropriate understanding comes from integrating out the fermionic degrees of freedom to determine the linear response.
We adopt this approach to confirm the equilibrium response of the system expected from $S_{CS}$. To $\mathcal{O}(e^2)$, after integrating out the fermions,
 \begin{eqnarray}\nonumber
    S&=&-e\int \frac{d^3qd\omega}{(2\pi)^4} \Tr\left[G_\lambda(\vec{q},\omega)\gamma^\mu\right]\tilde{A}^*_\mu(\vec{q},\omega)\\\label{PhotonAction}
    &&-\frac{e^2}{2}\int\frac{d^3qd\omega}{(2\pi)^4} \tilde{A}_{\mu}(\vec{q},\omega)\Pi_\lambda^{\mu\nu}(\vec{q},\omega)\tilde{A}^*_{\nu}(\vec{q},\omega) ,
 \end{eqnarray}
 where  $G_\lambda(\vec{q},\omega)$ is the single particle, interacting, Green's function in the presence of $B_z$ and $b_{z}$ and $\Pi_\lambda^{\mu\nu}(\vec{q},\omega)=\int \frac{d^3q'd\omega'}{(2\pi)^4}  \Tr\left[\gamma^\mu G_\lambda(\vec{q'},\omega')\gamma^\nu G_\lambda(\vec{q'}-\vec{q},\omega'-\omega)\right]$. The anomalous terms we are interested in can then be isolated by considering the leading $\vec{q},\omega\to 0$ terms which provide the static homogeneous response. 
 
The evaluation of $G_\lambda(\vec{q},\omega)$ cannot be carried out exactly  however we are only interested in computing the density response and the form of~\eqref{ChiralSeparationEffect} suggestive of an RPA approximation. Indeed, the low energy response in the longitudinal directions is determined solely by the LLL whose current and density responses are completely captured by an RPA summation owing to its reduced dimensionality. Using the non-interacting Green's function in the Landau level basis derived in~[\citenum{MiranskyShovkovy}] we obtain
\begin{equation}
    \lim_{\substack{\vec{q}\to 0\\ \omega\to 0}}\Pi^{\mu\nu}_\text{RPA}(\vec{q},\omega)=\left[\frac{1}{1+\lambda^2\frac{eB_z}{2\pi^2}}P_\parallel+P_{\perp} \right]^{\mu}_{\rho}\lim_{\substack{\vec{q}\to 0\\ \omega\to 0}}\Pi^{\rho\nu}_{0}(\vec{q},\omega) ,
\end{equation}
 where, for $\lambda^2_{\mu\nu}=\lambda^2\eta_{\mu\nu}$,  $P_\parallel=(1-\gamma_3)/2$ projects onto the longitudinal components, while $P_\perp=1-P_\parallel$ projects onto the transverse components.  When  $\lambda^2_{\mu\nu}=\lambda^2\eta_{0\mu}\eta_{0\nu}$ we use instead  $P_\parallel=[(1-\gamma_3)/2][(1-\gamma_5)/2]$ which projects only onto the temporal components.  We see here a screening of the density response due to the interactions while the transverse components are unaffected. The equilibrium Hall response is therefore the same as the free case, in agreement with~\eqref{HallResponse}.  The linear density response is then found after computing $\Lim{\vec{q}\to 0} \Lim{\omega\to 0} \Pi^{02}_0(\vec{q},\omega)/iq_x$. Surprisingly however, this vanishes. Thus the anomalous density response comes from the first term in~\eqref{PhotonAction} and can be attributed to the change in degeneracy of the LLL. The same RPA screening occurs for this term also and we find agreement with~\eqref{ChiralSeparationEffect}.

In the absence of $B_z$, the density response depends on all filled bands~\cite{ChenWuBurkov}. When it is present however, this is not the case and the density response is determined only by the LLL. Therefore  we can understand this by returning to our description of the LLL given in~\eqref{SymmetricBoson}.  The $S_b$ term can be accounted for by the inclusion of a chemical potential term $S_{S,b}=-\int d^2x \sqrt{N} b_z\partial_x\phi_S/\pi$. Recalling that $N=eB_z/2\pi$ is identified with the degeneracy of the LLL we compute the density response to $N\to N+\delta N$ and once again find agreement with~\eqref{ChiralSeparationEffect}. Furthermore, the modification of the anomalous terms is natural from this viewpoint as we can identify $(1+\lambda^2eB_z/2\pi^2)^{-1}$ as being the charge susceptibility  or the chiral charge stiffness of the LLL~\cite{GiamarchiBook, GogolinNerseyanTsvelikBook}. This is in agreement with \eqref{ChiralSeparationEffect} being viewed either as the density response or the CSE.



\textit{Conclusions}---
In this Letter we have explored the interplay between anomalous chiral symmetry breaking via electromagnetic fields and interactions. We have shown, using Fujikawa's path integral method, that the chiral charge continuity equation contains new interaction dependent terms which can be absorbed into effective electromagnetic fields which are responsible for the breaking of chiral symmetry. Furthermore this result was shown to be consistent with the lowest Landau level being a Luttinger liquid. We investigated the consequences of this result for interacting Weyl semimetals and found that interaction effects will be present in the non-equilibrium Hall response as well as the density response to a change in the magnetic field. These results were then reproduced via direct perturbative calculation.

 Recently, it was discovered that the circular photogalvanic effect~\cite{ deJuanGrushinMorimotoMoore}, originally thought to be quantized as a result of the chiral anomaly, is actually renormalized due to the presence of interactions~\cite{AvdoshkinKoziiMoore}. It would be desirable to understand our results in the context of this observable also. Lastly, we note that other anomalous Ward identities, including the gravitational anomaly can be derived using Fujikawa's method and our analysis can likewise be applied in those situations with the possibility of additional observable interaction effects~\cite{Gooth}.

\acknowledgements  We acknowledge useful discussions with Natan Andrei. This work was supported by the U.S. Department of Energy, Office of Science, Basic Energy Sciences under Award No. DE-SC0001911, the Simons Foundation (A.P., C.R., and V.G.)
and Natural Sciences and Engineering Research Council (NSERC) of Canada (AAB). 
Research at Perimeter Institute is supported in part by the Government of Canada through the Department of Innovation, Science and Economic Development and by the Province of Ontario through the Ministry of Economic Development, Job Creation and Trade.

\bibliography{mybib}

\end{document}